\documentclass[conference]{IEEEtran}
\IEEEoverridecommandlockouts
\usepackage{cite}
\usepackage{amsmath,amssymb,amsfonts}
\usepackage{algorithmic}
\usepackage{graphicx}
\usepackage{textcomp}
\usepackage{xcolor}
\def\BibTeX{{\rm B\kern-.05em{\sc i\kern-.025em b}\kern-.08em
    T\kern-.1667em\lower.7ex\hbox{E}\kern-.125emX}}
\begin{document}
\title{Portfolio Growth Rate and Ergodic Capacity of Fading Channels\\
\author{Cihan Tepedelenlioglu,  School of ECEE,  Arizona State University}
}
\maketitle
\renewcommand \baselinestretch{1}
\newcommand{\D}{\mathbf{D}}
\newcommand{\beq}{\begin{equation}}
\newcommand{\eeq}{\end{equation}}
\newcommand{\A}{\mathbf{A}}
\newtheorem{thm}{Theorem}
\def\LT{\leq_{LT}}
\def\c{\leq_{c}}
\def\G{\leq_{{\cal U}}}
\def\sG{{\cal U}}
\def\bb{{\bf b}}
\def\WX{W_{\bf X}}
\def\bt{{\tilde b}}
\def\bbt{{\tilde {\bf b}}}
\def\bX{{\bf X}}
\def\bXt{{\tilde {\bf X}}}
\def\xt{{\tilde x}}
\def\bY{{\bf Y}}
\def\bYt{{\tilde {\bf Y}}}
\def\bU{{\bf U}}
\def\bV{{\bf V}}
\def\bI{{\bf I}}
\def\bK{{\bf K}}
\def\bR{{\bf R}}
\def\bT{{\bf T}}
\def\a{\alpha}
\def\cX{{\cal X}}
\def\cY{{\cal Y}}
\def\be{{\bf e}}
\def\bH{{\bf H}}
\def\bx{{\bf x}}
\def\bv{{\bf v}}
\def\bS{{\bf \Sigma}}
\def\bL{{\bf \Lambda}}
\def\inc{\subseteq}
\begin{abstract}
A relationship between the growth-rate of log-optimal portfolios and capacity of fading single-input multiple output (SIMO) channels are established. 
Using this relation, stock vector stochastic processes that model the investment environments are stochastically ordered using different criteria. The presence of side information (SI) is considered, and a bound on the gains in the growth-rate due to SI is derived along with data processing inequality and convexity properties. A statistical test on the usefulness of SI that does not require the computation of the optimal portfolio vector in the presence of SI is introduced and its several variants are discussed.
\end{abstract}
\section{Introduction}
In portfolio theory, the goal is to maximize the combined stochastic portfolio value according to some utility function \cite{coverthomas}. It is well known that the optimization of the logarithmic utility function (also known as the Kelly criterion) leads to the best growth rate for constantly rebalanced portfolios, even though it does not always provide portfolios on the mean-variance efficiency frontier. Due to its constant-rebalancing feature, even when the individual stocks have zero growth rate, the combined portfolio value can exhibit positive growth rate through a so-called volatility pumping property  \cite{luenberger}. Log-optimal portfolios have long-term benefits such as achieving a large target wealth in the shortest time \cite{breiman}, but also have short-term optimality properties like maximizing the expected wealth-relative for one rebalancing period \cite{kelly,coverthomas}.

To recall the basic setup of log-optimal portfolio theory,  let $\bX=[X_0, \ldots, X_M]^T$ be a stock (column) vector consisting of nonnegative random variables according to a multivariate cumulative distribution function (CDF)  $F_\bX(\bx)$. Here, $X_m$ represent the price-relative of the $m^{th}$ stock which is the ratio of the price in two consecutive time intervals (rebalancing periods).  In log-optimal portfolio theory the goal is to maximize the growth rate given by 
\vspace{-.2 cm}
\begin{equation}
\vspace{-.3 cm}
\WX(\bb) = E[\log(\bb^T \bX)]=\int \log(\bb^T \bx) dF_\bX(\bx) d\bx \label{wb}
\end{equation}
where $\bb=[b_0 \ldots b_M]^T$ is the portfolio column vector which lives on the probability simplex ${\cal B}$ (with nonnegative elements which sum to one). This definition of the growth-rate is justified if the stock market stochastic process is  i.i.d. across time from the distribution $F_\bX(\bx)$, or more generally, it is a stationary and ergodic process with that CDF for each time slot. In that case, the portfolio vector $\bb$ that maximizes (\ref{wb}) also maximizes the growth-rate of the stochastic process 
\begin{equation}
S_N = \prod_{n=1}^{N} \bb^T \bX^{(n)} \label{sn}
\end{equation}
where $\bX^{(n)}$ is the realization of the stock vector in the $n^{th}$ rebalancing period.

For the special case of a horse-race, where the stock vector realization can only have a single nonzero entry, the optimization of (\ref{wb}) over the simplex yields a closed-form which is a decreasing function of the entropy rate of the random process $\bX^{(n)}$. In addition, asymptotic equipartitioning property for the stock market \cite{coverthomas} offers a connection between source coding and portfolio theory. 

The connection between portfolio optimization theory and information theory have mainly been in the relationship between the entropy-rate of the stock vector process and its growth rate. In this paper, a relationship between the growth rate in (\ref{wb}) to the capacity of a multi-antenna fading channel is established. This relationship is used to establish stochastic orders 
between two different stock vector distributions, which are partial orders on probability distributions, using stochastic orders  for fading channels \cite{rajan}. 

Our contributions in this paper are as follows:
\begin{itemize}
\item A relationship between the growth rate of a log-optimal portfolio and the ergodic capacity of single-input, multiple-output fading channels is established. Using this  connection, several stochastic orders between vector stock stochastic processes are defined, and these are related to stochastic orders defined for fading channels.
\item Information-theoretic properties of the financial value of information (FVSI) \cite{barron} which is the gain on the portfolio growth-rate due to the presence of side-information (SI) is derived. These include  convexity with respect to the transition kernel, and the data processing inequality. Also a new bound on the  FVSI is derived, which is only a property of the distribution of $\bX$.
\item A new statistical test for the usefulness of SI is introduced, which does not require  the   calculation of  the optimal portfolio in the presence of SI. The test is based on the Karush-Kuhn-Tucker (KKT) conditions for optimality. Different variants of this test are introduced including general utility functions.
\end{itemize}
%
\section{Growth Rate and Ergodic Capacity}
\subsection{Volatility and the Fractional Kelly Criterion}
Due to the volatility of the stochastic process $S_N$ in the short-term, it is often desirable to control its variability by blending it with a risk-free asset such as cash. Such a criterion is called a {\it fractional Kelly} portfolio where the fraction invested in the risk-free asset is fixed in advance.  Consider a stock market where a risk-free asset is available so that  $\bX^{(n)} = [1 \;\; \bXt^{(n)T}]^T$ or equivalently, the first entry of $\bX^{(n)}$ is $X^{(n)}_0=1$. When a fraction $\lambda$ is invested in cash, (\ref{sn}) becomes
\begin{equation}
S_N = \prod_{n=1}^{N} (\lambda + (1-\lambda) \bbt^T \bXt^{(n)}) \label{snl}
\end{equation}
where we are defining $\bb = [\lambda \;\; b_1 \ldots b_M]^T$ so that $b_0 =\lambda$, and $\bbt=[b_1 \ldots b_M]^T/(1-\lambda)$ which is on the probability simplex. 
Due to the fact that $\lambda=1$ yields zero variance for $S_N$, and because of the continuity of the variance of $S_N$ in $\lambda$, a desired value of variance can be set by adjusting $\lambda$ while  sacrificing growth rate.
Taking logs in (\ref{snl}) we obtain
\begin{equation}
\frac{1}{N} \log S_N = \frac{1}{N} \sum_{n=1}^{N} \log\left(1+ \frac{ 1-\lambda}{\lambda} \; \bbt^T \bXt^{(n)}\right) + \log(\lambda) \;. \label{logsn}
\end{equation}
It is clear from (\ref{logsn}) that for an ergodic stock process, maximizing the growth rate of $S_N$ with respect to $\bbt$  for a fixed $\lambda$ amounts to maximizing 
\begin{equation}
E\left[\log\left(1+ \frac{1-\lambda}{\lambda}\; \bbt^T \bXt\right)\right] \label{pcap} 
\end{equation}
with respect to $\bbt$ over the simplex, where we dropped the time index due to the stationarity of the stock vector process.
If the goal is the best possible growth rate, then the fraction $\lambda$ can also be optimized. However, if a large growth rate subject to a desired variance is sought, then the $\lambda$ parameter will be selected to satisfy that constraint. 
\subsection{Ergodic Capacity of Fading Channels}
To draw a parallel between the growth-rate of the value of a portfolio and the ergodic capacity of fading channels, consider a single-input, multi-output channel model 
\cite{alouini,tse}:
\begin{equation}
y_m^{(n)} = \sqrt{\rho_m} h_m^{(n)} s^{(n)} + v_m^{(n)} , \;\; m=1,\ldots, M,
\end{equation}
where $\rho_m$ is the nonnegative power allocated on the $m^{th}$ branch with combined power 
$\rho := \sum_m \rho_m$; $h_m^{(n)}$ is the complex-valued channel gain, $s^{(n)}$ is the transmitted symbol, and $v_m^{(n)}$ is the complex Gaussian additive noise at time $n$. The channels, input symbol and the noise samples all have zero-mean and normalized variance of unity.  We consider a setup where the transmitter does not know the channel coefficients but the receiver does. 

For a diversity combining receiver with complex-valued combining coefficients $\alpha_m$, the signal to noise ratio (SNR)-maximizing choice is given by $\alpha_m = \sqrt{\rho_m} h_m^*$, where $(\cdot)^*$ denotes complex conjugate, and we drop the time index for convenience. The resulting combined instantaneous SNR is given by 
\begin{equation}
{\rm SNR} = \sum_{m=1}^M \rho_m |h_m|^2 \;.
\end{equation}
If the channel coefficients $h_m^{(n)}$ are stationary across the time index $n$, the Shannon capacity of this channel is given by \cite{tse}
\begin{equation}
C = E\left[\log\left(1+\sum_{m=1}^M \rho_m |h_m|^2\right)\right]\;. \label{cap}
\end{equation}
Note that (\ref{cap}) and (\ref{pcap}) have the same functional form with the identification $\bbt^T=[\rho_1 \ldots \rho_M]/\rho$, $\bXt^T = [|h_1|^2 \ldots |h_M|^2]$, and due to the normalization of the channel input, channel coefficient and noise variance, $\rho=(1-\lambda)/\lambda$ can be interpreted as the average SNR. Therefore, finding the optimal $\lambda-$ fractional Kelly portfolio $\bbt$ is identical to finding the optimal power allocation over the diversity branches. This establishes a link between growth-rate optimal portfolios, and ergodic capacity of fading channels. 
The similarity in (\ref{cap}) and (\ref{pcap}) will enable us to extend stochastic orders for fading channels to those in portfolio theory.
\subsection{Water Filling}
Before moving to stochastic orders, we want to continue pointing out the mathematical similarity between fractional Kelly portfolios and capacity maximization in communication theory by showing that (\ref{pcap}) can be reduced to the celebrated water-filling problem \cite{tse} if the stock vector is a horse race. To see this, consider
the scenario where $\bXt$ can only take a single nonzero entry for every realization with equal probability:  $P[\bXt = \be_m \xt_m]=1/M$. 
Here $\be_m$ is the unit vector in the $m^{th}$ direction. Defining ${\tilde b}_m = b_m/(1-\lambda)$ we see that (\ref{pcap}) reduces to 
\begin{eqnarray}
\frac{1}{M} \sum_{m=1}^M \log \left(1+\frac{1-\lambda}{\lambda}\; {\tilde b}_m \xt_m \right) \;. \label{wf}
\end{eqnarray}
Ignoring the constant $1/M$ factor, the optimization of (\ref{wf}) with respect to $\bbt$ over the simplex for a given set of $\xt_m$ is precisely the same problem 
of optimizing the capacity of parallel channels and a total power constraint. This solution (see e.g., \cite{coverthomas}) has the well-known  water-filling interpretation.
\section{Stochastic Orders}
Stochastic orders (or dominance) are partial orders on probability distributions. Our interest here will be to order two stock market environments using different criteria. In \cite{dominance} the authors considered an overview of stochastic dominance for portfolios with emphasis on first and second order stochastic dominance. Reference \cite{levystodominance} considered extensions to the multi-period case ($N>1$). In \cite{denuit} Laplace transform ordering is considered and its properties are developed. In \cite{rajan} a stochastic order based on the Shannon transform is developed for ordering random communication channel models. In what follows, we will introduce and relate these orders and study their implications on the ordering of growth rate of portfolios.

The stochastic orders discussed here will all be integral stochastic orders defined as follows. Let $\sG$ be a class of real-valued functions and $X$ and $Y$ be random variables. We define a generic integral stochastic order as 
\begin{eqnarray}
X \G Y \Longleftrightarrow E[u(X)] \leq E[u(Y)] \;, \forall u \in \sG \;, \label{generic}
\end{eqnarray}
where $\sG$ is called the generator function set of the stochastic order. Note that a given stochastic order might have more than one generator. When $\sG$ is the set of all increasing functions, we have the so-called usual stochastic order, which can be shown to be equivalent to point-wise ordering of the CDFs of $X$ and $Y$. If, in addition to being increasing, the functions in $\sG$ are also concave, then we have second-order stochastic order in (\ref{generic}). Continuing in this manner, one can require more strict stochastic orders by adding the requirement that $\sG$ contain functions whose derivatives alternate in sign. This naturally leads to the Laplace transform ordering defined next.
\subsection{Laplace Transform Order}
When $\sG = \{u(x) = -\exp(-\rho x), \rho \geq 0\}$ we have 
\begin{eqnarray}
X \LT Y \Leftrightarrow E[\exp(-\rho Y)] \leq E[\exp(-\rho X)]\; ,  \rho \geq 0 \;. \label{lt}
\end{eqnarray}
The same Laplace transform order can also be defined by adopting $\sG$ as the set of all functions with a completely monotone derivative. 
These are functions whose derivatives are completely monotone (c.m.), where c.m. functions are  positive functions with alternating derivatives:
$$(-1)^n d^nu(x)/dx^n \geq 0 \;, \forall x >0,  \; n =0,1,2,\ldots$$ 
\subsection{Ergodic Capacity Order}
The LT order can be coarsened by considering $\sG$ which is a proper subset of functions with a completely monotone derivative. In particular, 
 $\sG =\{u(x) = \log(1+\rho x), \rho \geq 0\}$ can be used to define an integral stochastic order: 
 \begin{eqnarray}
X \c Y \Leftrightarrow E[\log(1+\rho X)] \leq E[\log(1+\rho Y)]\; ,  \rho \geq 0 \;. \label{lt}
\end{eqnarray}
 This stochastic order, which is defined through the so-called Shannon transform of the distribution of a positive random variable, was originally defined in \cite{rajan}
 in the context of fading communication channels. There it was shown that LT ordering implies ergodic capacity ordering, but the converse does not hold. 
 \subsection{Stochastic Ordering of Stock Vector Processes}
 The different integral stochastic orders defined above provide a framework for comparing the scalar portfolio values which are random variables. There are several ways to extend this to vector-valued stock processes. 
We will now use these stochastic orders to define stochastic orders on stock market vector processes, and relate them to orders that can be defined through the growth rate. Toward this goal, consider the following orderings:
\begin{subequations}
\label{grecolto}
\begin{align}
E[\log \bb^T \bX] & \leq E[\log \bb^T \bY] \;, \forall \bb \in {\cal B}  \label{gr} \\
\bb^T \bX &\c  \bb^T \bY \;, \forall \bb \in {\cal B} \label{eco} \\
\bb^T \bX &\LT  \bb^T \bY\;, \forall \bb \in {\cal B}  \label{lto}
\end{align}
\end{subequations}
where ${\cal B}$ is the probability simplex.

Because Laplace transform ordering implies ergodic capacity ordering, we have (\ref{lto}) $\Rightarrow$ (\ref{eco}).
Moreover, because $\log$ has a c.m. derivative, we have  (\ref{lto}) $\Rightarrow$ (\ref{gr}). Because (\ref{gr}) indicates that the stock market environments $\bX$ and $\bY$ are such that their growth rates will be ordered for every portfolio selection, the LT ordering of stock vector processes
implies that their optimal growth rates are also ordered. 

One can also establish a partial connection between (\ref{gr}) and (\ref{eco}). Indeed, if (\ref{gr}) holds, then using (\ref{logsn}) and (\ref{pcap})
we have
$$
E\left[\log\left(1+ \frac{1-\lambda}{\lambda}\; \bbt^T \bXt\right)\right] \leq E\left[\log\left(1+ \frac{1-\lambda}{\lambda}\; \bbt^T \bYt\right)\right] \;, \label{pcapo} 
$$
which holds for all $\bbt \in {\cal B}$ and $0 \leq \lambda \leq 1$. In other words, ordering of the growth rates of stock vectors $\bX=[1 \; \bXt]$ and $\bY=[1 \; \bYt]$ in the presence of a cash option $(X_0=1)$ implies the ergodic capacity ordering of the risky assetts $\bXt$, and $\bYt$.

The stock vector orderings in (\ref{grecolto}) have the downside of being dependent on the order of elements within the vectors $\bX$ and $\bY$. It is possible to impose permutation invariance by inserting a permutation matrix on the left-hand side of  equations  (\ref{grecolto}) and requiring them to hold for every permutation matrix. While satisfying permutation-invariance, this would have the down-side of making these stochastic orders too stringent in that it would require the worst stock on the right hand side to be better than the best one on the left-hand side if $\bb$ is chosen as a unit vector, for example. 

In an effort to impose less conditions to establish an order, consider the following alternative to (\ref{grecolto}):
\begin{subequations}
\small
\label{grecolto2}
\begin{align}
\max_{\bb \in {\cal B}} E[\log \bb^T \bX] & \leq  \max_{\bb \in {\cal B}} E[\log \bb^T \bY] \;,   \label{gr2} \\
\max_{\bb \in {\cal B}} E[\log(1+\rho \bb^T \bX)] &\leq  \max_{\bb \in {\cal B}} E[\log(1+\rho \bb^T \bY)]\; , \forall \rho \geq 0 \label{eco2}  \\
\max_{\bb \in {\cal B}} E[\exp(-\rho \bb^T \bY)] &\leq \max_{\bb \in {\cal B}} E[\exp(-\rho \bb^T \bX)]\; ,  \forall \rho \geq 0 \;. \label{lto2}
\end{align}
\end{subequations}
Note that in the $\bb$ values on either side of the inequality are optimized (and different), in all three equations in (\ref{grecolto2}).
This also makes these orders permutation-invariant.
It is clear that (\ref{gr2}) is simply comparing the growth rates of the two stock vectors under their respective optimal portfolios. This constitutes a total order. The order in (\ref{eco2}) compares two stock vector processes on the basis of their optimal fractional Kelly portfolio values for every fraction $\lambda$ (note the correspondence $\rho = (1-\lambda)/\lambda$ in (\ref{pcap})).  In (\ref{lto2}) since the optimal values of $\bb$ used on either side of the equation are dependent on $\rho$, it cannot be written as the LT ordering of two random variables.
\section{Side Information}
Given a stock market vector process $\bX^{(n)}$, if a SI stochastic process $S^{(n)}$ is available, then the portfolio selection can benefit from the knowledge of realizations of the SI so that the optimization of (\ref{wb}) can be performed over portfolio-valued functions $\bb(\cdot)$ where $\bb(s) \in {\cal B}, \forall s$. We will assume that $(\bX^{(n)},S^{(n)})$ have a joint distribution given by $p_{\bX,S}(\bx,s)$. The amount of improvement in the growth rate from using the side information is termed ``financial value of information'' (FVSI) in \cite{barron} and defined as
\begin{equation}
V(\bX;S) := \max_{\bb(\cdot)} E[\log \bb^T(S) \bX] - \max_{\bb} E[\log \bb^T \bX] \;, \label{V}
\end{equation}
where the first maximization is with respect to every portfolio-valued function $b(\cdot)$ of $S$, while the second one is over vectors in ${\cal B}$. Clearly, $V(\bX;S) \geq 0$ and in \cite{barron} it was shown to be upper bounded by the mutual information: 
\vspace{-.2 cm}
\beq
\vspace{-.3 cm}
V(\bX;S) \leq I(\bX;S) \label{mib}
\eeq
with equality for a horse-race market. 

In the context of SIMO fading channels, if a SI random variable is available at the transmitter side about the channel realizations, then the improvement in channel capacity can be bounded similar to (\ref{mib}) using the mutual information between the channel coefficients and SI. The results that follow are presented in a portfolio optimization context, but they all have analogues for fading channels.
\subsection{Bounding the FVSI}
Consider a motivating example where $\bX=[X_0 \; X_1] = [1 \; X_1]$ where $X_0=1$ with probability one and $X_1$ is a discrete random variable that satisfies $X_1 \geq 1$ with probability 1, so that $X_1$ is always the best stock. Let $H(X_1)>0$ the entropy of $X_1$, be strictly positive. Here the optimal portfolio will always pick $X_1$ so that  $\bb\strut^\star = [0 \; 1]^T$. No side information random variable in this setting will offer any gains in the growth rate. In particular, if $S=X_1$ we have 
$0 = V(\bX;S) < I(\bX;S) = H(X_1)$. A similar example to this was provided in \cite{barron} to illustrate that the mutual information bound on the FVSI can be loose. The reason this bound is loose is because the distribution of $\bX$ is such that the best stock always has the same index. We now provide a simple bound on the FVSI that is given by the entropy of the random index of the best stock. 

Define $m^\star = {\rm argmax}_m = X_m$ as the (random) index of the best stock. Since the knowledge of $m^\star$ will always enable choosing the best stock, it is the best possible SI random variable in terms of the FVSI: 
\vspace{-.1 cm}
$$
\vspace{-.2 cm}
V(\bX;S) \leq V(\bX;m^\star) \leq I(\bX;m^\star)=H(m^\star)-H(m^\star|\bX) \;.$$ Since 
$m^\star$ is a function of $\bX$, $H(m^\star|\bX) =0$ and  we have the following:
$V(\bX;S) \leq H(m^\star) \;. $
Especially in cases where $I(\bX;S)$ can be large (e.g., when $\bX$ and $S$ are continuous), this bound can be more useful than the mutual information bound. Note that the bound only depends on the stock distribution, and not the conditional distribution $p_{S|\bX}$.
\vspace{-.1 cm}
\subsection{Properties of SI}
\subsubsection{Convexity in $p_{S|\bX}$} To see that $V(\bX;S)$ is convex in $p_{S|\bX}$ for a fixed $p_S$, we observe that (\ref{V}) only depends on  $p_{S|\bX}$ through the first term on the right hand side given by
\begin{equation}
\max_{\bb(\cdot)}  \int p_\bX(\bx) \int p_{S|\bX}(s|\bx) \log(\bb^T(s) \bx) ds d\bx
\end{equation}
which is the maximum of a linear function of $p_{S|\bX}$, and therefore convex. 
\subsubsection{Data Processing Inequality (DPI)} If $\bX - S - Z$ is a Markov chain, then $V(\bX;S) \geq V(\bX;Z)$. 
To see this, we write $V(\bX;S) - V(\bX;Z)$ using (\ref{V})  as
\begin{equation}
\max_{\bb(\cdot)} E[\log \bb^T(S) \bX]  - \max_{\bb(\cdot)} E[\log \bb^T(Z) \bX] \;.
\label{tw}
\end{equation}
Since $\bX - S - Z$, given the random variable $S$, it is possible to generate a realization of $Z$ using the conditional distribution $p_{Z|S}$ and use this $Z$ value in the portfolio-valued function that optimizes 
$E[\log \bb^T(Z) \bX]$. This will be a randomized portfolio vector which has the same average utility
as $\max_{\bb(\cdot)} E[\log \bb^T(Z) \bX]$. The presence of such a randomized portfolio can be used to infer that there is a deterministic portfolio with that same performance. This shows that the difference in (\ref{tw}) is nonnegative.
\subsection{A Consistent Test for the Usefulness of SI}
We now develop a statistical test for the financial relevance of side information. Clearly, if the side information is statistically independent of the stock vector, it cannot provide any benefit. However, the converse is not true: when the side information is statistically dependent of the stock vector, it still might provide no benefit. So statistical tests for independence do not provide a full solution for the problem of testing for FVSI. 
In \cite{mathissi} the authors used a universal portfolio framework \cite{universalsi,coveruniversal} where the stock sequence was deterministic to 
test for side information. The computation of the test statistic in this setting  requires the calculation of the constant optimal portfolio in the absence of SI,  and the piece-wise constant optimal portfolio in the presence of SI. The ratio of the value of the two portfolios was used as a test statistic. Here we take a more conventional approach by assuming a random stock sequence, and the test statistic does not require the computation of the optimal portfolio in the presence of SI.

The KKT conditions for the log-optimal portfolio optimization problem (maximizing (\ref{wb}) over the simplex) is given by \cite[pp. 617]{coverthomas}
\begin{equation}
\small
E\left[\frac{X_m}{\bb\strut^{\star T} \bX}\right] = 1 \label{kkt} 
\end{equation}
for the optimal solution $\bb\strut^\star$ and  the {\it active stocks} $m \in {\cal A}$ which are the stocks for which  $b_m^\star >0$. When $m \notin {\cal A}$,  $b^\star_m=0$, and the KKT conditions require only $\leq 1$ in (\ref{kkt}). These are necessary and sufficient conditions for optimality due to the concavity of the objective function. Let us define the side information random variable $S$ as having a finite range over the set $\{1,2, \ldots, K\}$. If, instead, $S$ is a continuous random variable, then the test can be applied to any of its quantizations (see \cite{erkip} for the rate distortion theory of quantizing SI). Due to the data processing inequality, if any quantization of SI reveals the presence of FVSI, then that means the SI has financial relevance. 

Applying the KKT conditions in (\ref{kkt}) to the conditional distribution of $\bX^{(n)}$ given $S$, we can obtain  necessary and sufficient condition for the side information vector to provide no benefit: 
\vspace{-.1 cm}\begin{equation}
\vspace{-.3 cm}
\small
E\left[\frac{X_m}{\bb\strut^{\star T} \bX}\; \bigg|\;S=k\right] = 1, \;\;\;
m \in {\cal A},  \; k \in \{1, \ldots, K\}
 \label{kktside}
\end{equation}
where $\bb\strut^\star$ is still the optimal portfolio in the absence of side information.

\subsubsection{Test Statistic}
Suppose we have access to stock vector and side information realizations $(\bX^{(n)},S^{(n)})$ for $n=1,\ldots,N$. This test is for a setting where the optimal portfolio without side information has already been acquired, and we are seeking to know if SI information would offer any benefit, without the need to compute the conditionally optimal portfolio in detail. 

We now construct a test for the usefulness of side information by testing the null hypothesis given in (\ref{kktside}) which asserts that the unconditionally optimal portfolio $\bb^\star$ remains optimal for every state $k$ of the side information $S$. The optimal portfolio $\bb^\star$ in the absence of SI can be computed by using e.g. \cite{coveralg} to optimize (\ref{wb}) where the expectation can be approximated using a sample average over $\bX^{(n)}$.  So if  $\bb^\star$ is available, then whether the side information is useful or not can be tested by estimating the left hand side of (\ref{kktside}) and checking how close it is to the right hand side by constructing the following test statistic: 
\begin{equation}
T:=\sum_{m \in {\cal A}} \sum_{k=1}^K \left|\frac{1}{N_k}\sum_{n\in {\cal I}_k} \frac{X^{(n)}_m}{\bb\strut^{\star T} \bX^{(n)}} -1 \right|^2 \;, \label{test}
\end{equation}
where ${\cal I}_k$ is the set of all indices $n$ such that $S^{(n)}=k$, and $N_k$ is the number of elements in ${\cal I}_k$, so that $\sum_k N_k =N$. Due to the law of large numbers  (LLN) $T \rightarrow 0$ as $N\rightarrow \infty$ with probability one. However, how fast this occurs depends on the underlying distribution.  This is because the null hypothesis that ``the side information is not useful" is a composite hypothesis. In other words, when (\ref{kktside}) holds, this does not specify the distribution of $(\bX^{(n)},S^{(n)})$. Therefore, for any threshold $\tau$, the probability $P[T > \tau]$ is a function of the joint distribution of the random variables
$(\bX^{(n)},S^{(n)})$. 
\subsubsection{Testing Individual Stocks}
It is possible to refine the test statistic in (\ref{test}) to determine which values of the side information $k$  affect which stocks $m$. To see this, define $T_{m,k}$ as the squared absolute value term in (\ref{test}) so that $T=\sum_{m,k} T_{m,k}$. By individually testing the $K(M+1)$  null hypotheses $T_{m,k} >\tau_{m,k}$, one can  determine if ``the $k^{th}$ side information $S=k$ is useful for the $m^{th}$ stock".  These  tests are still consistent for each $(m,k)$ due to the LLN. Moreover, their false-alarm probabilities for each $m,k$ can be approximated using the central-limit theorem.
\subsubsection{Short Selling}
If short selling is allowed, then one no longer has the constraint $b_m \geq 0$ while retaining $\sum_m b_m =1$ (see e.g. \cite{luenberger}). In the absence of this positivity constraint, the optimality conditions are always given by equality in (\ref{kkt}) and (\ref{kktside}) for all stocks, including those for which $b^\star_m=0$. In this case, the test statistic is simply given as in (\ref{test}) with the modification that the sum over $m$ is over $\{0,\ldots,M\}$.
\subsubsection{General Utility Functions}
Instead of the $\log$ utility in (\ref{wb}), consider a general strictly concave increasing utility function $u(\cdot)$. In this case the KKT conditions in (\ref{kktside}) become
\begin{equation}
\small \hspace{-.15 cm}
E\hspace{-1 mm} \left[X_m u'(\bb\strut^{\star T} \bX)\; \bigg|\;\hspace{-1 mm}S=k\right] = 
E\hspace{-1 mm}\left[(\bb\strut^{\star T} \bX) u'(\bb\strut^{\star T} \bX)\; \bigg|\;S=k\right] 
\label{kktsidegen}
\end{equation}
for $m \in {\cal A},  \; k \in \{1, \ldots, K\}$, where $u'(\cdot)$ is the first derivative of $u(\cdot)$.
From the point of view of testing for the FVSI the only substantive difference with (\ref{test}) is that the right-hand-side of (\ref{kktsidegen}) is not necessarily a constant 1. This requires also estimating the right-hand-side as part of the test statistic:
\begin{equation}
\small
T_{\rm gen}:=\sum_{m \in {\cal A}} \sum_{k=1}^K \left|\frac{1}{N_k}\sum_{n\in {\cal I}_k} 
u'(\bb\strut^{\star T} \bX^{(n)})\left(X_m^{(n)}  - \bb\strut^{\star T} \bX^{(n)}\right)
 \right|^2. \label{testgen}
\end{equation}
For example, $u(x) = (x^{\alpha}-1)/\alpha$ for $0 < \alpha \leq 1$ is a (power) utility function with a completely-monotone derivative (and therefore is a concave increasing function). In this case $u'(x) = x^{\alpha-1}$.
For $\alpha$ very close to zero, this utility function behaves very similar to a $\log$ utility function, in which case $T_{\rm gen}$ and $T$ are identical.
 \subsubsection{Test Performance}
 The test in (\ref{test}) is consistent for any ergodic stock sequence in the sense that $T$ converges to zero as $N\rightarrow \infty$. However, to set an appropriate threshold, one needs to determine the distribution of (\ref{test}) under the null hypothesis. Moreover, a bound on the false alarm probability that does not depend on the individual distributions is desirable. As a first step toward this goal, we will assume that $S^{(n)}$ and $\bX^{(n)}$ are independent of each other, and i.i.d. across time under the null hypothesis. In this case, it can be shown that $T$ is asymptotically a sum of squares of Gaussians where each term has a variance that is upper-bounded by 
\begin{equation}
\theta =  \frac{1}{N\min_k P[S=k]} \left(\frac{1}{\min_{m\in {\cal A}} b_m^2}-1 \right)
\end{equation}
and the false alarm probability can be bounded by
 \begin{equation*}
P[T > \tau |H_0]  \leq 1-\frac{1}{\Gamma((M+1)K/2)}\gamma\left((M+1)K/2,\tau/\theta \right) \;,
 \end{equation*}
 where $\gamma(\cdot,\cdot)$ is the lower incomplete gamma function.
A detailed analysis of the false-alarm and missed detection probabilities of the tests developed is possible in the asymptotic regime (large $N$) even when $S$ and $\bX$ are not independent. In that case,  the false-alarm probability depends on the (potentially unknown) distribution of the composite null hypothesis, and therefore it is desirable to find upper bounds that are distribution-independent.
\section{Conclusion and Future Work}
A link between growth-rate of fractional Kelly portfolios and ergodic capacity of SIMO channels is established. Vector extensions of stochastic orders based on several different utility functions are considered, including one that naturally arises in ergodic capacity of fading channels. These can also be extended to continuous geometric Brownian motion models. The convexity and DPI properties of the SI can be further exploited in gaining a better understanding of quantization of SI. Unlike conventional quantization, in portfolio optimization, quantization of SI is necessary not because it is convenient for its communication, but also for computational reasons which make the calculation of the optimal portfolio-valued function feasible. In future work, different variants of FVSI tests will be considered and asymptotic analysis of false alarm and detection probabilities will be carried out.
\newpage
\bibliographystyle{IEEEtran}
\bibliography{portfoliorefs}
\end{document}